# Photoproduction of long-lived holes and electronic processes in intrinsic electric fields seen through photoinduced absorption and dichroism in $Ca_3Ga_{2-x}Mn_xGe_3O_{12}$ garnets

V V Eremenko[1], S L Gnatchenko[1], I S Kachur[1], V G Piryatinskaya[1], A M Ratner[1], M B Kosmyna[2], B P Nazarenko[2] and V M Puzikov[2]

[1] Institute for Low Temperature Physics and Engineering, National Academy of Sciences of Ukraine, 61103 Kharkov, Ukraine
[2] Scientific-Research Department of Optical and Constructional Crystals, STC 'Institute for Single Crystals', National Academy of Sciences of Ukraine, 61001 Kharkov, Ukraine

E-mail: Piryatinskaya@ilt.kharkov.ua

**Abstract**
Long-lived photoinduced absorption and dichroism in the $Ca_3Ga_{2-x}Mn_xGe_3O_{12}$ garnets with $x < 0.06$ were examined versus temperature and pumping intensity. Unusual features of the kinetics of photoinduced phenomena are indicative of the underlying electronic processes. The comparison with the case of $Ca_3Mn_2Ge_3O_{12}$, explored earlier by the authors, permits one to finally establish the main common mechanisms of photoinduced absorption and dichroism caused by random electric fields of photoproduced charges (hole polarons). The rate of their diffusion and relaxation through recombination is strongly influenced by the same fields, whose large statistical straggling is responsible for a broad continuous set of relaxation components (observed in the relaxation time range from 1 to about 1000 min). For $Ca_3Ga_{2-x}Mn_xGe_3O_{12}$, the time and temperature dependences of photoinduced absorption and dichroism bear a strong imprint of structure imperfection increasing with $x$.

## 1. Introduction and statement of problem

Long-lived photoinduced phenomena in magnetic insulators were observed by many research groups [1–13] and associated with the photoinduced transfer of charges. In ferromagnets or ferrimagnets, whose Curie temperature exceeds the upper temperature boundary of the existence of long-lived photoinduced phenomena, interrelated photoinduced changes in optical and magnetic properties make up a complicated physical picture difficult for microscopic interpretation. For example, the illumination of yttrium iron garnet $Y_3Fe_5O_{12}$ with linearly polarized light induces a change of magnetic anisotropy [1–3], linear dichroism [3] and domain structure [4]. Unpolarized light affects magnetic permeability and susceptibility [5, 6], coercivity and mobility of domain walls [7, 8], magnetostriction [9] and optical absorption [10, 11].

To explore the microscopic mechanism of long-lived photoinduced phenomena, it is helpful to separate photoinduced changes of optical properties from those of magnetic characteristics related to a magnetically ordered state. To that end, the authors [12, 13] recently examined the antiferromagnetic garnet $Ca_3Mn_2Ge_3O_{12}$ within a broad temperature region between the Néel temperature $T_N = 13.85$ K and the temperature boundary (at about 170 K) of the existing long-lived changes. The absence of magnetic ordering in this temperature region simplified photoinduced phenomena and enabled us to elucidate their nature in the following way.

We observed in the $Ca_3Mn_2Ge_3O_{12}$ garnet long-lived photoinduced absorption and dichroism which exhibited unusual spectral and kinetic features. The spectrum of the photoinduced addition to absorption was found to be near to an initial absorption band whereas the decay kinetics of the photoinduced absorption and dichroism consisted of a continuous set of components with decay times varying over a range of three orders of magnitude. As temperature rises, the observed decay kinetics does not noticeably change while the photoinduced addition to absorption diminishes by an order of magnitude. The observed set of decay components cannot be ascribed to some new irradiation-produced optical centres but is naturally explained by random electric fields of photoproduced localized charges (long-lived holes are capable of hopping over the lattice whereas the compensating negative background is formed by pinned charges). These fields manifest themselves in two ways: first, they enhance a forbidden optical transition and the corresponding absorption band; second, they strongly accelerate the recombination of holes with negative charges, a broad continuous set of decay times being conditioned by a continuous distribution of the random fields over magnitude.

Thus, it was shown that photoinduced absorption (its spectrum, relaxation kinetics, temperature dependence, dichroism) provides information on photoproduced holes and the action of their electric fields. Based on this information, several questions should be answered:

(1) In what way do electric fields of photoproduced holes enhance the oscillator strength of the optical transition serving as a probe for these fields?
(2) In what way do pumping photons, being absorbed within a specific absorption band, create long-lived holes?
(3) What is the mechanism of the relaxation (disappearance) of photoproduced holes?
(4) What properties of holes, produced by linearly polarized light, carry information on the polarization direction and are responsible for dichroism?

In [12, 13], these questions were partially answered as applied to the garnet $Ca_3Mn_2Ge_3O_{12}$:

(1) The transition, enhanced by the electric field of photoproduced holes, was identified as the forbidden transition between even states $^5E_g$ and $^5T_{2g}$. The forbiddenness is partially eliminated by an electric field that imparts an odd addition to even states.
(2) The photoproduction of charges was associated with a small impurity of $Mn^{4+}$ ions in the $Mn^{3+}$ sublattice. In the ground state, these Mn holes are pinned near Ge vacancies bearing a compensating negative charge; light pumping of the Mn sublattice results in a partial separation of these positive and negative charges and in the corresponding enhancement of their electric field.

(3) The relaxation of photoproduced holes, i.e. their recombination with negative charges, occurs via hopping of hole polarons with overcoming an energy barrier that is lowered by the total electric field of all photoproduced charges. The statistic straggling of these fields causes a straggling of the relaxation rate by several orders of magnitude. As the temperature rises, this broad distribution of holes over relaxation rates shifts to the right, resulting in the observed temperature evolution of the relaxation kinetics (see section 4).

To obtain a more detailed and definite answer to the above questions, it is helpful to trace how the change of the medium, where photoproduced holes move, and the corresponding change of optical transitions, used for pumping and probing, will affect the entire picture observed.

In the present work, bearing in mind this purpose, we examine the garnets $Ca_3Ga_{2-x}Mn_xGe_3O_{12}$ with $x < 0.06$ in order to compare the character of photoinduced phenomena throughout a set of garnets including the case of $x = 2$ (i.e. the $Ca_3Mn_2Ge_3O_{12}$ garnet) and the cases of $x \approx 0.055, 0.02, 0.01$. Such a comparison is informative due to the following essential distinctions between the cases of $x = 2$ and $x < 0.06$:

(a) In the interval $x \leqslant 0.02$, where an absorption spectrum in the visible range is formed mainly by $Mn^{4+}$ ions, the optical spectrum of $Ca_3Ga_{2-x}Mn_xGe_3O_{12}$ is quite different from that of the $Ca_3Mn_2Ge_3O_{12}$ garnet with a regular $Mn^{3+}$ sublattice [14] (see section 3); hence, the absorption bands used for pumping and the probe are also different. This difference is helpful in tracing both the way of photoproduction of long-lived holes and the mechanism of enhancement of a probe optical transition by their electric fields.

(b) Unlike the garnet $Ca_3Mn_2Ge_3O_{12}$ with the periodic $Mn^{3+}$ sublattice, a crystal $Ca_3Ga_{2-x}Mn_xGe_3O_{12}$ with $x < 0.06$ contains manganese in the form of spatially isolated ions, which must essentially impede the motion of Mn holes and the relaxation of photoproduced charges associated with Mn holes. Hence, the absence of a strong $x$ dependence of the character of the relaxation kinetics [14] (see section 4) gives the grounds to associate the photoproduced charges with another type of hole moving in a regular sublattice. Simultaneously, this corroborates the general relaxation scheme developed earlier [12, 13], irrespective of the specific nature of the photoproduced charged polarons.

## 2. Experimental details

Single crystals of the $Ca_3Ga_{2-x}Mn_xGe_3O_{12}$ garnet were grown by the Czochralski method in platinum crucibles with 50 mm diameter and 60 mm height in an air–argon atmosphere using inductive heating. Necessary thermal conditions in the crystal growth zone were provided by an additional platinum screen heater whose geometry and position could be varied and fitted within the crucible.

The crystal was pulled from the melt at a rate of 2 mm h$^{-1}$, the rotation rate being in the interval of 30–40 rotations per min. The initial charge was prepared from the mixture of $GeO_2$, $Ga_2O_3$, $MnO_2$ and $CaCO_3$ by the solid-phase synthesis technique. The stability of the crystal growth regime and the suppression of the $GeO_2$ evaporation from the melt were achieved by small temperature gradients in the crystal and melt and by an optimal choice of the excess gas pressure (1–1.5 atm). The grown crystals had a high structural quality and uniform colouration, and were free of blocks and inclusions. The size of the obtained single crystals was 25 mm in diameter and 50–70 mm in height.

The samples were cut in the form of plane-parallel plates perpendicular to the [100] direction. The thickness of the plates was varied in the interval 0.05–1.5 mm in order to make the sample sufficiently transparent within the spectral region to be examined.

Manganese was introduced into the charge of $Ca_3Ga_{2-x}Mn_xGe_3O_{12}$ in the content $x_{melt} = 0.02$, 0.1 and 0.5. The corresponding content of Mn in the crystals was estimated by optical methods (see section 3) as 0.01, 0.02 and 0.055, respectively.

The measurements of absorption spectra of these samples were carried out with the use of a diffractional monochromator with linear dispersion of 1.3 nm mm$^{-1}$. The study of photoinduced phenomena was carried out with an optical double-beam set-up. A sample was illuminated by green light with the wavelength at the maximum of intensity $\lambda_{max} = 510$ nm, spectral halfwidth of 50 nm and flux density of 0.14 W cm$^{-2}$. As a source of pumping, a helium–neon laser with a wavelength 633 nm was also used (the laser flux density was 0.13 W cm$^{-2}$). A stable wide-band emission of a xenon arc lamp, dispersed through the monochromator, served as the probe light. The intensity of the probe beam was low enough to cause no photoinduced phenomena. After passing through the sample, the probe beam was run through a second monochromator tuned to the wavelength of the probe light in order to cut off the scattered pumping light. After that, the probe beam was registered by a photomultiplier. The electric signal from the photomultiplier was amplified, transformed to digital form by an analogue–digital converter and transmitted to a computer.

For the measurements of photoinduced dichroism, pumping light was polarized in the direction [110] of the crystal (this polarization direction provides the strongest photoinduced dichroism). The probe light, after passing through the sample, was split by a polarizer into two beams polarized in mutually perpendicular directions ([110] and [$\bar{1}$10]). Then a rotating disc opened the path for these beams in turn for equal time intervals. After passing through the second monochromator, both beams were in turn registered by the same photomultiplier. The detailed description of the optical set-up is given in [13].

## 3. Absorption spectra: estimation of the content of manganese in crystals

Figure 1 presents absorption spectra of the $Ca_3Ga_{2-x}Mn_xGe_3O_{12}$ samples grown from melt with different manganese content: $x_{melt} = 0.02, 0.1, 0.5$. For comparison, the low-frequency wing of the absorption band of the $Ca_3Mn_2Ge_3O_{12}$ garnet [12] is shown.

The analysis of these absorption spectra can be carried out on the basis of the following considerations. As is known [15], impurity manganese can enter the lattice of the $Ca_3Ga_2Ge_3O_{12}$ garnet in the form of $Mn^{2+}$, $Mn^{3+}$ and $Mn^{4+}$ ions which occupy different positions in the unit cell. The $Mn^{2+}$ ions substitute $Ca^{2+}$ at the dodecahedral positions whereas $Mn^{3+}$ ions substitute $Ga^{3+}$ at the octahedral positions. $Mn^{4+}$ ions also enter octahedral positions instead of $Ga^{3+}$, an excess positive charge being compensated for either by $Ge^{4+}$ vacancies or via the substitution of $Ge^{4+}$ ions by $Ga^{3+}$ in a corresponding number of sites [15]. However, these mechanisms can compensate only for a small excess charge, so that the concentration of $Mn^{4+}$ in the lattice rapidly saturates with an increase of the concentration of manganese in melt.

Analysis of absorption spectra of the garnets $Ca_3Ga_{2-x}Mn_xGe_3O_{12}$ with different concentrations of manganese at a temperature $T = 32$ K (figure 1) shows that, at small $x$, impurity manganese contributes to the absorption spectrum mainly in the form of $Mn^{4+}$ ions. A predominant contribution of $Mn^{4+}$ to absorption can be distinctly traced for the $Ca_3Ga_{2-x}Mn_xGe_3O_{12}$ sample 2 grown from the melt with manganese content $x_{melt} = 0.1$ (curve 2 in figure 1). This spectrum contains a broad band with the maximum at 520 nm, as well as a low-frequency wing of another broad band which is very strong in the ultraviolet region. These absorption bands are typical of compounds containing $Mn^{4+}$ in the octahedral surroundings of oxygen ions [16–19]. The absorption band with a maximum at 520 nm was established to correspond to the orbitally forbidden optical transition $^4A_2 \rightarrow {}^4T_2$ ($^4F$) in the $Mn^{4+}$ ion placed in the crystalline field of $O_h$ symmetry; the position of this band in figure 1 practically coincides with that observed in [17] in the garnet $Ca_3Ga_2Ge_3O_{12}$:$Mn^{4+}$. The intensive absorption band in the ultraviolet region (whose wing is better seen in figure 2) was ascribed to charge transfer between the adjacent ions $O^{2-}$ and $Mn^{4+}$ [16, 19]. Another orbitally forbidden transition in $Mn^{4+}$ $^4A_2 \rightarrow {}^4T_1$ ($^4F$) gives rise to a weak absorption band lying in the region 350–380 nm according to the calculation data of [16, 18, 19]. This band weakly manifests itself near 370 nm on a strong background of the mentioned charge-transfer band (figure 2).

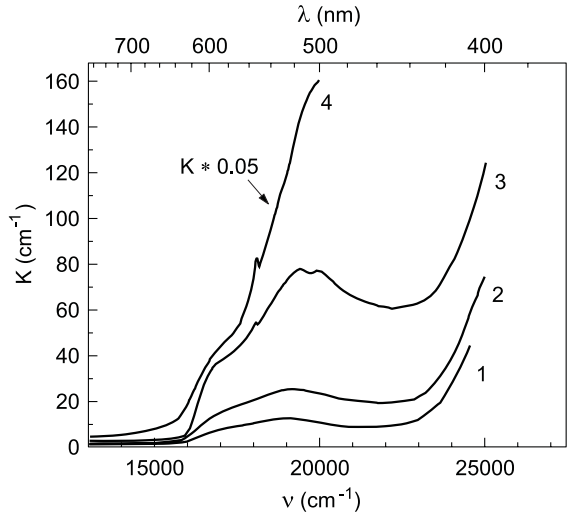

**Figure 1.** Absorption spectra of samples of $Ca_3Ga_{2-x}Mn_xGe_3O_{12}$ with various $x$ measured at 32 K: 1—$x = 0.01$, 2—$x = 0.02$, 3—$x = 0.055$. For comparison, the absorption spectrum of $Ca_3Mn_2Ge_3O_{12}$ at 32 K [12] is presented (curve 4).

**Table 1.** Estimation of the content, $x = x_3 + x_4$, of manganese in samples of $Ca_3Ga_{2-x}Mn_xGe_3O_{12}$ from the absorption spectrum ($x_3$ and $x_4$ are the contributions of $Mn^{3+}$ and $Mn^{4+}$, respectively).

| Sample number | Mn content in melt, $x_{melt}$ | $Mn^{4+}$ content in crystal, $x_4$ | $Mn^{3+}$ content in crystal, $x_3$ | Total Mn content in crystal, $x$ | $x/x_{melt}$ |
|---|---|---|---|---|---|
| 1 | 0.02 | 0.01 | Not detected spectroscopically | 0.01 | 0.5 |
| 2 | 0.1 | 0.02 | spectroscopically | 0.02 | 0.2 |
| 3 | 0.5 | 0.035 | 0.02 | 0.055 | 0.11 |

Taking the aforesaid into account, the content, $x$, of manganese in a $Ca_3Ga_{2-x}Mn_xGe_3O_{12}$ sample can be estimated from the absorption spectrum as the sum

$$x = x_4 + x_3, \qquad (1)$$

of the contributions of $Mn^{4+}$ and $Mn^{3+}$ ions, respectively.

The content, $x_4$, of $Mn^{4+}$ in a $Ca_3Ga_{2-x}Mn_xGe_3O_{12}$ sample can be estimated via comparison of the absorption coefficient at 520 nm within the same band $^4A_2 \rightarrow {}^4T_2$ ($^4F$) for the sample examined and for a crystal $YAlO_3$:Mn with a known concentration of $Mn^{4+}$ investigated in [20]. The estimates, obtained in this way for three samples of $Ca_3Ga_{2-x}Mn_xGe_3O_{12}$, are given in table 1.

$Mn^{3+}$ ions, lying in octahedral sites, usually manifest themselves in the absorption spectrum by a broad band with a maximum near 500 nm, corresponding to the orbitally forbidden transition $^5E \to {}^5T_2$ ($^5D$) [21, 22]. In $Ca_3Mn_2Ge_3O_{12}$, only a low-frequency tail of this band can be observed because of a strong absorption at its maximum (see figure 1, curve 4). Of all the examined $Ca_3Ga_{2-x}Mn_xGe_3O_{12}$ samples, only in sample 3 ($x_{melt} = 0.5$, $x \approx 0.055$) does this band manifest itself as an additional maximum at 500 nm. The content of $Mn^{3+}$ in sample 3 was estimated as $x_3 \approx 0.02$ via evaluating the absorption intensity in this additional maximum and comparing it with the data on absorption of various garnets [22]. The total content, $x = x_3 + x_4$, of manganese in the crystal is presented in the last but one column of table 1. The last column shows the portion of manganese in the melt that enters the lattice of the grown crystal. The content of manganese dissolved in the crystal saturates with an increase of $x_{melt}$ at the level of about $x = 0.06$; it is impossible to grow a crystal with a greater $x$ of a satisfactory structural quality using this method.

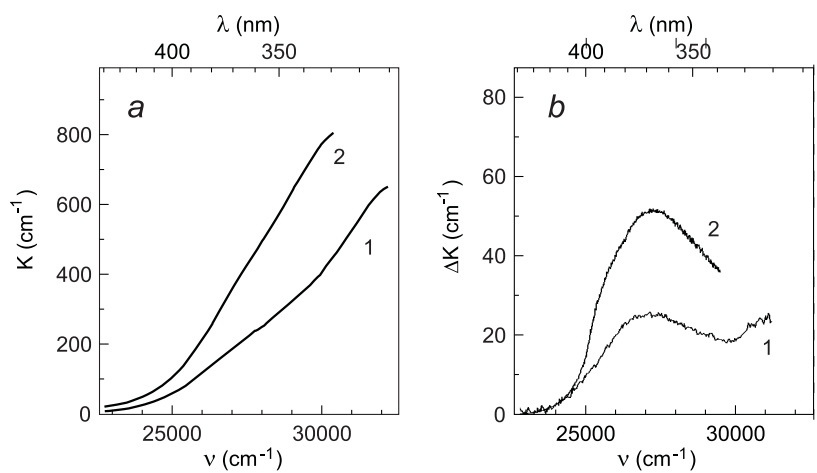

**Figure 2.** Absorption spectra (a) and spectra of photoinduced addition to absorption (b) at 32 K in the region of short waves where photoinduced absorption is significant. 1—$x = 0.01$, 2—$x = 0.02$. The samples were irradiated with green light (pumping intensity has a maximum at $\lambda_{pum} \approx 510$ nm).

It cannot be excluded that some portion of manganese also enters dodecahedral sites as $Mn^{2+}$ ions. However, $Mn^{2+}$ ions can hardly be discovered optically within the spectral range examined; indeed, all optical transitions from their ground state are forbidden both orbitally and by spin, which makes their manifestation in absorption indiscernible on the background of a much stronger absorption of $Mn^{3+}$ and $Mn^{4+}$ ions. We have not discovered any band typical of $Mn^{2+}$ ions in the absorption spectra of the examined samples.

## 4. Photoinduced absorption, its kinetics and nature

### 4.1. Experimental data

Photoillumination of crystals $Ca_3Ga_{2-x}Mn_xGe_3O_{12}$ results in an enhancement of optical absorption in the wavelength region $\lambda < 440$ nm. Figure 2 presents the spectrum of the augmentation, $\Delta K$, of the absorption coefficient caused by the illumination with green light ($\lambda_{max} = 510$ nm) at $T = 32$ K. Under illumination with an He–Ne laser ($\lambda = 633$ nm), a similar effect with the same spectral dependence of $\Delta K$ was observed but $\Delta K$ was about three times less in magnitude (with rising temperature this difference diminishes).

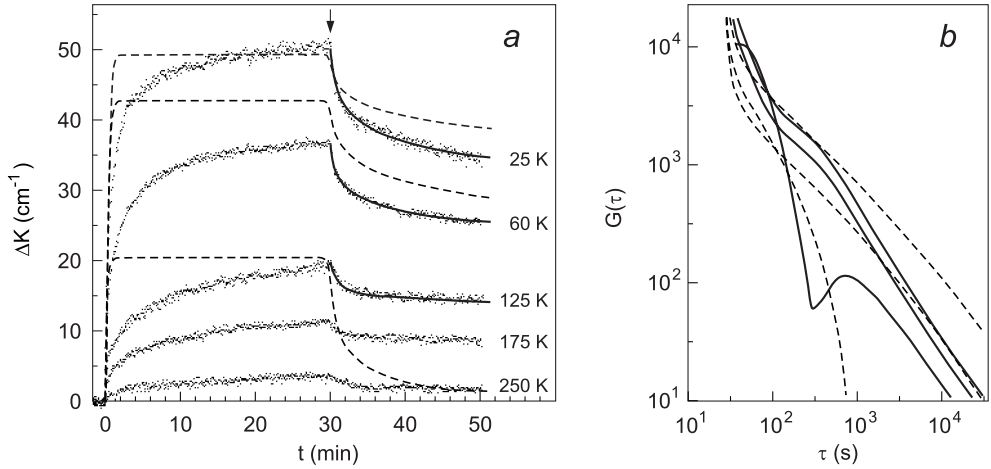

**Figure 3.** (a) Time dependences of the photoinduced addition to absorption for the sample with $x = 0.02$ measured at various temperatures (indicated in the figure) under full pumping. An arrow indicates the moment of switching off the irradiation. Full curves show the relaxation kinetics in the form of expansion (2) with the function $G(\tau)$ derived from experiment and shown in figure 3(b). Broken curves present the solution of kinetic equation (5). (b) Full curves: distribution, $G(\tau)$, of decay components over decay times derived by the best fitting of expansion (2) with experiment for $T = 25$, 60 and 125 K (from top to bottom). Broken curves show the lifetime distribution of holes in random fields according to (3)–(4) for $T = 25$, 60 and 125 K (from top to bottom). The difference of the curves (3)–(4) from the experimental curve at $T = 125$ K is indicative of structural inhomogeneities, as shown in section 6.

Photoinduced absorption $\Delta K$ depends on manganese content $x$ in a nonmonotonic way (which is due to a structural imperfection growing with $x$, see section 6, item 1). As is seen from figure 2, $\Delta K$ increases two times as $x$ increases from 0.01 (sample 1) to 0.02 (sample 2). With a further increase of $x$, $\Delta K$ diminishes.

Figure 3(a) shows the time dependence of $\Delta K$ under pumping ($t \leqslant 30$ min) and during the subsequent relaxation without pumping ($30 < t < 50$ min), measured for $x = 0.02$ at various temperatures. Although photoinduced absorption is thermally suppressed above 200 K, an increase of temperature does not accelerate the relaxation of $\Delta K$ after switching off illumination.

Figure 4 shows by full squares the temperature dependence of photoinduced absorption $\Delta K$ for $x \approx 0.02$ (sample 2) achieved after half an hour of pumping in a regime near to saturation. In the case of the $Ca_3Mn_2Ge_3O_{12}$ garnet, presented in the figure for comparison, $\Delta K$ diminishes with temperature significantly faster (which will be explained in section 6, item 2).

### 4.2. Nature of photoinduced absorption

The kinetics of photoinduced absorption gives a key to the nature of photoinduced phenomena. First, let us focus on the relaxation kinetics in order to base the choice between two possible mechanisms of photoinduced absorption. The latter can be *a priori* attributed to:

(a) the creation of some new absorption centres;
(b) the photoinduced enhancement of an absorption band that exists irrespective of pumping. This means that the oscillator strength of the corresponding transition in absorption centres is enhanced by the electric field of photoproduced charges.

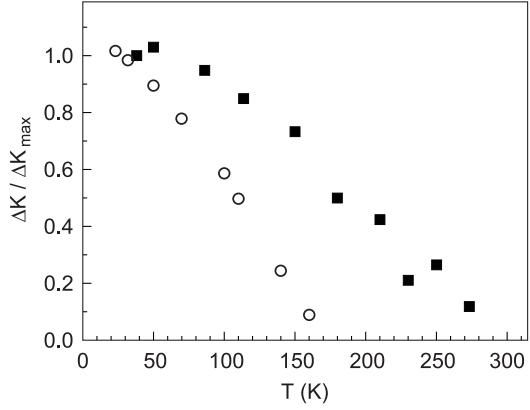

**Figure 4.** Squares: temperature dependence of photoinduced absorption $\Delta K$ for $x = 0.02$ registered at $\lambda_{reg} = 420$ nm under pumping with red light ($\lambda_{pum} = 633$ nm). For comparison, a similar dependence is shown for $Ca_3Mn_2Ge_3O_{12}$ ($\lambda_{pum} = 633$ nm, $\lambda_{reg} = 565$ nm) by the circles [13]. Photoinduced absorption $\Delta K$ is expressed in units of its maximum value.

As is seen from figure 3(a), the relaxation kinetics of $\Delta K$ after switching off illumination consists of a set of components with decay times $\tau$ varying in the range of one minute to many hours:

$$\Delta K(t) = \int G(\tau) \exp(-t/\tau) \, d\tau. \tag{2}$$

The weights, $G(\tau)$, of all $\tau$ components in this expansion were obtained via fitting equation (2) to the observed relaxation kinetics (figure 3(a)) with the use of the standard linear least-square method. The derived $\tau$ distribution $G(\tau)$ for $T = 25, 60$ and $125$ K is presented in figure 3(b) in the double logarithmic scale; expansion (2) with this function $G(\tau)$ is plotted in figure 3(a) by the full curve.

Figure 3(b) shows that the mentioned interval of $\tau$ from 0.5 min to several hours is filled in continuously by the decay components (for $T = 125$ K, a decrease at $t \approx 300$ s is due to structural inhomogeneities, as will be shown in section 6, item 2). On the other hand, as is seen immediately from figure 3(a), both short-$\tau$ and long-$\tau$ components are substantially present in the relaxation kinetics. Thus, the contribution of the entire $\tau$ interval to the relaxation kinetics is essential (although $G(\tau)$ rapidly decreases with $\tau$, long-$\tau$ components are integrated in (2) over a much larger interval than short-$\tau$ components).

It is difficult to describe such relaxation kinetics by a discrete set of photoproduced absorption centres of different types: this would require a large number of types and their broad smooth $\tau$ distribution. Version (a) cannot also explain other experimental facts adduced at the end of this section. Thus, version (a) should be discarded.

The mechanism (b), described below, naturally eliminates the above difficulty. The relaxation rate $\Gamma = 1/\tau$ of a photoproduced hole is dictated by the rate of its thermally activated hopping, strongly influenced by the electric field, $\boldsymbol{F}$, of all charges [13]:

$$\frac{1}{\tau} \equiv \Gamma = \Gamma_0 \exp\left(-\frac{Q - e\boldsymbol{F}\boldsymbol{a}}{2T_{eff}}\right) \tag{3}$$

($T_{eff}$ differs from the true temperature $T$ by making allowance for the zero lattice vibrations at low temperatures). Equation (3) describes the hopping of a hole with a charge $e$ by a distance $\boldsymbol{a}$ to an adjacent site in the direction of the applied field $\boldsymbol{F}$. This field lowers the hopping activation energy $Q$ by $e\boldsymbol{F}\boldsymbol{a}$. The expression (3) for hopping probability (in the main obvious) was derived in [13] through the analysis of statistic fluctuations around the initial and final sites of the hopping (their simultaneous allowance resulted in the appearance of the multiplier 2 in the denominator of the exponent).

The random electric field $F$ at the hole position point is the sum of the statistically independent electric fields of all other holes. The distribution of such a sum over its magnitude is generally described by a Gaussian that, in our case, assumes a form [13]

$$P(F) = \text{constant} \times F^2 \exp(-0.0215(\varepsilon/e)^2 N^{-4/3} F^2) \quad (4)$$

with effective permittivity $\varepsilon$ and the number of holes per unit volume $N$. The magnitude distribution (4) of random fields has a large dispersion, which provides a very broad continuous distribution of holes over lifetime (3). The lifetime distribution of holes, described by equations (3) and (4), is shown in figure 3(b) by broken curves; at low temperatures, it mainly coincides with the expansion of the experimental decay curve over $\tau$ components. Thus, the straggling of photoinduced fields over magnitudes naturally explains the presence of a broad set of $\tau$ components in the relaxation kinetics observed.

In the kinetic equation [12, 13], the $\tau$ distribution of holes (3) and (4) is taken into account via dividing them into $M$ groups numbered by subscript $j$. The lifetime, $\tau_j$, of the $j$th group is influenced by a field $F_j$, according to (3). The nodes $F_j$ are separated by intervals $\Delta F_j$ proportional to $1/P(F_j)$, so that a photoproduced hole gets with the same probability, $1/M$, into every interval $\Delta F_j$. The distribution of holes over such groups corresponds to their distribution over lifetimes, $G(\tau)$, in expansion (2).

Making an allowance for this, the population, $n_j$, of the $j$th group is described by a kinetic equation [12, 13]

$$\frac{\mathrm{d}n_j}{\mathrm{d}t} = \frac{J}{M} - \hat{\Gamma} n_j. \quad (5)$$

Here the first term describes the production of a hole by pumping ($J$ is its intensity duely normalized). The second term describes a complicated relaxation process which has a different character under pumping and in the absence of pumping. In the latter case, the second term in (5) simply turns into $n_j/\tau_j$. In the former case, pumping creates new holes and simultaneously accelerates the relaxation of already existing holes in the following way. During a time space $t_{pum} \propto 1/J$ (of the order of 1 min), pumping creates a number of new holes comparable with that of the existing holes, entailing an essential change of the random fields at the position points of all holes. As a result, during the time $t_{pum}$ every hole gets, with a probability of about 0.5, into a strong field and disappears, so that the actual lifetime of a hole is limited by $t_{pum}$:

$$\tau_j^{\mathit{eff}} = \min\{\tau_j, t_{pum} \propto 1/J\}. \quad (6)$$

This means that the hole lifetime under pumping strongly differs (by more than two orders of magnitude) from that in the absence of pumping. In the kinetic equation (5), this relaxation process is allowed for by the operator $\Gamma$ explicitly disclosed in [13].

The kinetic equation (5) was solved numerically with nearly the same parameters as in [12, 13] (the maximal achievable concentration of photoproduced holes was put equal to 1 mol%). Figure 3(a) presents by broken curves the calculated photoinduced absorption, written as $AN \equiv A\Sigma n_j$ and fitted to experiment by the scale multiplier $A$. The theory describes satisfactorily the relaxation of $\Delta K$, observed after switching off illumination at a low temperature, but significantly overestimates the relaxation rate at high temperatures. This discrepancy is caused by a structural inhomogeneity of the $Ca_3Ga_{2-x}Mn_xGe_3O_{12}$ samples (see section 6, item 2) and is much weaker for the garnet $Ca_3Mn_2Ge_3O_{12}$ with an almost homogeneous structure [12, 13].

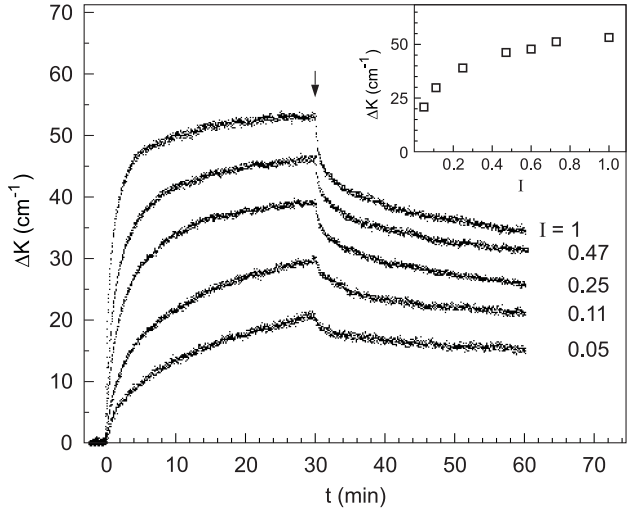

**Figure 5.** Time dependences of the photoinduced addition to absorption for the sample with $x = 0.02$ ($\lambda_{pum} \approx 510$ nm, $\lambda_{reg} = 375$ nm, $T = 32$ K) under pumping with various intensities (indicated in the figure in units of the maximal intensity). The arrow indicates the moment of switching off irradiation. Inset: pumping dependence of photoinduced absorption measured after 30 min irradiation of the sample.

The time dependence of $\Delta K$, observed under pumping after its switching on, essentially deviates from the theoretical curve by the time required to achieve the saturation level $\Delta K_{sat}$. Table 2 characterizes this discrepancy by the time, $t_{0.8}$, during which photoinduced absorption $\Delta K$ achieves the value $0.8 \Delta K_{sat}$. This discrepancy is strong for the $Ca_3Ga_{2-x}Mn_xGe_3O_{12}$ samples with noticeable structural inhomogeneities but weak for the $Ca_3Mn_2Ge_3O_{12}$ garnet with almost perfect structure. The role of the structural inhomogeneities will be elucidated in section 6, item 2.

Here let us point out a feature of kinetics inherent in the homogeneous case: the time required to achieve saturation of $\Delta K$ under pumping ($\sim 1$ min) essentially differs from the characteristic relaxation time in the absence of pumping ($\geqslant 1$ h). This difference, naturally explained by equation (6) within the model of photoproduced charges, cannot be understood within the usual kinetic equation $dN/dt = J - N/\tau$, that describes the processes of saturation and relaxation with the same relaxation time $\tau$.

Note also that the usual kinetic equation, in a drastic contradiction with experiment, gives the saturation value $\tau J$ proportional to pumping intensity $J$. Contrary to this, equation (6) gives the saturation value $\Delta K_{sat} = \tau^{eff} J$ almost independent of $J$. This inference is confirmed by the experimental time dependence of $\Delta K$ observed at $T = 32$ K under pumping with different intensities (figure 5). Table 2 presents the ratio of two values of $\Delta K_{sat}$ achieved under full pumping ($I = 1$) and at $I = 0.05$; this ratio, derived from the kinetic equation (5), equals 1.25, which does not strongly deviate from experiment (especially for $Ca_3Mn_2Ge_3O_{12}$ with almost homogeneous structure).

## 5. Nature of photoproduced charges

In accordance with the consideration of section 3, the photoinduced enhancement of absorption, as well as in the case of the $Ca_3Mn_2Ge_3O_{12}$ garnet [12, 13], is caused by the electric field of the photoproduced charges. However, an essential difference exists. In the $Ca_3Mn_2Ge_3O_{12}$ garnet, the photoinduced field enhances the *orbitally forbidden* $^5E_g \rightarrow {}^5T_{2g}$ transition in every $Mn^{3+}$ ion of the Mn sublattice. This forbiddenness is partially eliminated by a photoinduced field $F \sim 10^6$ V cm$^{-1}$ weak compared to the intra-atomic field $F_0 \sim 10^8$ V cm$^{-1}$. The photoinduced field causes an addition to the absorption coefficient $\Delta K \sim K_0(F/F_0)^2 \sim 100$ cm$^{-1}$, where $K_0 \sim 10^6$ cm$^{-1}$ is the absorption coefficient of an allowed transition. In the case of $Ca_3Ga_{2-x}Mn_xGe_3O_{12}$, an *allowed* transition (with charge transfer between $O^{2-}$ and $Mn^{4+}$) seems to be less sensitive to an applied field; besides, $Mn^{4+}$ ions are present in a low concentration ($x \approx 0.02$). Nevertheless, $\Delta K$ is found to be of the same order of magnitude.

**Table 2.** The time, $t_{0.8}$, required to achieve the level of $0.8\Delta K_{sat}$ after switching on the full pumping; the ratio of two saturation values $\Delta K_{sat}(I)$ related to different values of pumping $I = 1$ and $0.05$ ($T = 32$ K). First line: the results of the calculation with the use of equation (5); second and third lines: experimental data for $Ca_3Mn_2Ge_3O_{12}$ and $Ca_3Ga_{2-x}Mn_xGe_3O_{12}$ with $x \approx 0.02$.

|  | $t_{0.8}$ (min) | $\Delta K_{sat}(1)/\Delta K_{sat}(0.05)$ |
|---|---|---|
| Theory (equation (5)) | 0.6 | 1.25 |
| $Ca_3Mn_2Ge_3O_{12}$ | 0.7 | 1.3 |
| $Ca_3Ga_{2-x}Mn_xGe_3O_{12}$ | 2.5 | 2 |

To understand this surprising experimental result, it is necessary to consider the nature of photoproduced charges. In the case of $Ca_3Mn_2Ge_3O_{12}$ [12, 13], we associated them with a small number of Mn holes, i.e. with $Mn^{4+}$ ions whose charge was compensated on average by germanium vacancies bearing a negative effective charge. In the ground state, Mn holes are pinned at the Mn sites adjacent to germanium vacancies. It was assumed that the excitation of the $Mn^{3+}$ sublattice with pumping light results in de-pinned Mn holes moving in a polaronic state over the Mn sublattice.

For the $Ca_3Ga_{2-x}Mn_xGe_3O_{12}$ garnets with $x < 0.06$, such a mechanism is not applicable in view of the absence of the Mn sublattice. Let us consider possible types of photoproduced charges which arise after the excitation of a Mn ion (most probably, a $Mn^{4+}$ one). When creating a pair of opposite charges at the expense of the photon energy of 2.5 eV, an electron can be taken away only from an $O^{2-}$ ion. Indeed, ionization potentials of other lattice ions exceed 50 eV, whereas the creation of an oxygen hole requires a much lower energy (roughly 2 eV, judging from the electron affinity of a free $O^{2-}$ ion or from the position of the oxygen hole band in $YBa_2Cu_3O_6$ [23]). Thus, the photoproduction of a long-lived charge involves the creation of a hole in the $O^{2-}$ sublattice near a $Mn^{4+}$ ion turning to $Mn^{3+}$.

In more detail, the photoproduction of long-lived charges must be realized through the following stages.

(1) A pumping photon brings a $Mn^{4+}$ ion into an excited state. If the pumping light is linearly polarized, the excited state of the Mn ion is polarized correspondingly.
(2) The excitation energy of the $Mn^{4+}$ ion is expended in creating a free hole in the oxygen sublattice, an electron being completely transferred from $O^{2-}$ to the excited $Mn^{4+}$ ion (below called a Mn acceptor). This is possible if the threshold energy, $E_{th}$, required for creating an oxygen hole (of the order of 2 eV) does not exceed the energy, $E_{exc}$, of the initial atomic excitation. The difference:

$$E_{kin} = E_{exc} - E_{th}, \qquad (7)$$

turns into the kinetic energy of the created hole. Such an electronic process (a hole analogue of autoionization) requires overcoming some energy barrier that determines the lifetime, $\tau_{exc}$, of the $Mn^{4+}$ ion in the excited state (the time of its radiative decay is assumed to exceed $\tau_{exc}$). After overcoming this barrier, irrespective of $\tau_{exc}$, two opposite charges appear: an oxygen hole and an effective negative charge of the populated $Mn^{4+}$ acceptor (turned into $Mn^{3+}$). However, a long $\tau_{exc}$ promotes the relaxation of dichroism induced by polarized pumping.

(3) The oxygen hole, created in a free state, rapidly covers in the $O^{2-}$ sublattice a distance $l$, exceeding several lattice periods, before coming into a polaronic state with a low mobility. This is necessary to sufficiently separate the positive and negative charges (otherwise, they will recombine immediately).

(4) A hole in the oxygen sublattice comes to a two-site polaronic state well known in solid neon ($O^{2-}$ and Ne have the identical closed-shell configuration $1s^2 2s^2 2p^6$). The existence of two-site oxygen holes, important for explaining dichroism (see section 7), is shown at the end of this section.

(5) As was mentioned, such a two-site hole polaron is formed far enough from that excited $Mn^{4+}$ ion that had given rise to it, being turned into $Mn^{3+}$ (or, in short, a Mn acceptor). Therefore, a hole polaron, in the course of its diffusion over the oxygen sublattice, will most probably meet one of the unexcited $Mn^{4+}$ ions (strongly prevailing in number over the populated Mn acceptors) and get to a potential well of a dipole formed by the $Mn^{4+}$ ion and the compensating negative charge. In such potential wells hole polarons spend most of their lifetime.

(6) The relaxation of the photoinduced effect consists in the recombination of hole polarons with populated $Mn^{4+}$ acceptors (such photoproduced $Mn^{3+}$ ions, positioned near compensating negative charges, attract hole polarons stronger than the usual $Mn^{3+}$ ions). The recombination of a hole polaron is preceded by its random shift between potential wells associated with $Mn^{4+}$ ions.

Based on the above assumptions, it is easy to estimate the photoinduced addition, $\Delta K$, to the absorption coefficient related to the charge-transfer transition in $Mn^{4+}$ ions. The photoinduced effect is conditioned by the energy shift of the charge-transfer transition made by the electric field of photoproduced charges. The main contribution to the field at the centre of a $Mn^{4+}$ ion is made by a two-site hole localized in its first coordination sphere (figure 6). This energy shift is different for various versions of the charge-transfer transition related to different $O^{2-}$ ions of the first coordination sphere. Let us place at its centre the pole of spherical coordinates (figure 6). For an $O^{2-}$ ion with spherical angle $\theta$, the energy shift to the red is

$$\Delta E(\theta) \approx Fa \cos \chi = \frac{e}{\varepsilon r^2} a \cos \chi = \frac{1}{4\varepsilon a \sin(\theta/2)}. \tag{8}$$

Here $F$ is the electric field of the hole at the centre, A, of the $O^{2-}$ ion, $a = 0.2$ nm is the radius of the first coordination sphere, $r$ is the vector through point A and hole centre C, $\varepsilon \approx 2.5$ stands for an effective permittivity [13] and angles $\theta$ and $\chi = (\pi - \theta)/2$ are shown in figure 6. Averaging equation (8) over $\theta$ gives an estimate for the negative energy shift of the charge-transfer transition:

$$\Delta E \approx e/2\varepsilon a \approx 1 \text{ eV}. \tag{9}$$

This corresponds to the shift of the charge-transfer absorption band by about 8000 cm$^{-1}$ to the red. As seen from figure 2 (related to $x = 0.02$), such a shift would enhance the absorption coefficient by about 1000 cm$^{-1}$ supposing every $Mn^{4+}$ ion has a hole in its first coordination sphere. The real addition to the absorption coefficient can be roughly estimated as

$$\Delta K \sim 1000 \zeta \text{ cm}^{-1} \qquad \text{for } x = 0.02 \tag{10}$$

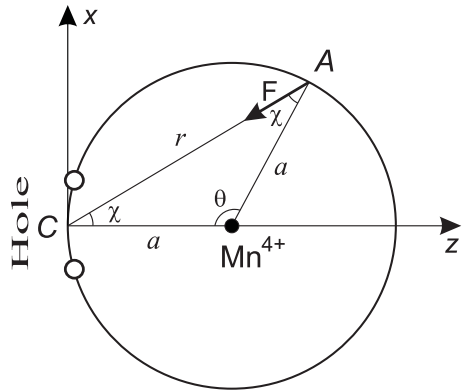

**Figure 6.** Two-site oxygen hole (small circles) trapped in the first coordination sphere of a $Mn^{4+}$ ion. The coordination sphere (large circle) is occupied by $O^{2+}$ ions, one of which is labelled by A.

where $\zeta$ is the portion of $Mn^{4+}$ ions transferring holes to the oxygen sublattice under pumping. As seen from figure 2, $\Delta K$ achieves 50 cm$^{-1}$ at the frequency 27 000 cm$^{-1}$; hence, $\zeta \sim 0.05$, i.e. about 5% of $Mn^{4+}$ ions turn into $Mn^{3+}$.

The following seeming contradiction should be pointed out. In the case of $Ca_3Mn_2Ge_3O_{12}$, nearly the same number of photoproduced holes caused an addition to the absorption of the same order of magnitude [13] although the number of probe optical centres ($Mn^{3+}$ ions) was about 100 times greater. The point is that every hole strongly affects only that probe centre with which it comes into immediate contact (gets to an adjacent site). So, the number of probe centres strongly affected is determined by the number of holes, no matter what the total number of probe centres is (assuming it exceeds the number of holes).

In conclusion, let us consider the structure of a hole polaron in the sublattice of the $O^{2-}$ ions with the closed-shell configuration of neon $1s^2 2s^2 2p^6$. The structure of the stable state of a hole polaron is well known for crystals of Ne, Ar, Kr and Xe, formed by atoms with a closed valence shell $ns^2np^6$. The valence $n$p hole, distributed between two adjacent atoms separated by a distance $r$, realizes a strong exchange binding (of the scale of 1 eV) that sharply increases with decreasing $r$ and makes two adjacent atoms strongly approach one another [24]. Similar two-site holes (called $V_k$ centres), formed in the anion sublattice of alkali halides, are also well known as a stable polaronic state (see, e.g., [25, 26]). In all cases, the exchange binding mainly depends on the ratio $\rho/b$, where $\rho$ is the valence p-state radius and $b$ is the corresponding interatomic distance in the ideal lattice. This ratio is given in table 3 for the $O^{2-}$ sublattice of the $Ca_3Ga_2Ge_3O_{12}$ garnet and, for comparison, for some of the crystals mentioned. Since the $\rho/b$ ratio for the $O^{2-}$ sublattice of the garnet is not less than for other crystals, where the existence of two-atom hole polarons has been reliably established, they must exist also in the $O^{2-}$ sublattice.

## 6. Influence of structural inhomogeneities on photoinduced absorption

Structural inhomogeneities of the $Ca_3Ga_{2-x}Mn_xGe_3O_{12}$ garnet, caused by the Mn impurity with a limited solubility, play a substantial role in photoinduced processes and manifest themselves in what follows.

**Table 3.** Valence p-state radius, $\rho$, and the interatomic distance in the ideal lattice, $b$, whose ratio determines the interatomic exchange interaction, forming a two-site hole polaron in some crystals (Å).

| Crystal | $\rho$ (Å) | $b$ (Å) | $\rho/b$ |
|---|---|---|---|
| $Ca_3Ga_2Ge_3O_{12}$ | 0.81 | 2.78 | 0.291 |
| Ne | 0.52 | 3.16 | 0.165 |
| Ar | 0.92 | 3.76 | 0.245 |
| Kr | 1.06 | 3.99 | 0.266 |
| KCl | 0.86 | 4.45 | 0.193 |
| KI | 1.25 | 4.98 | 0.251 |
| NaI | 1.25 | 4.57 | 0.273 |

*6.1. Nonmonotonic x dependence of photoinduced absorption $\Delta K$.*

This phenomenon, stated at the beginning of section 4, is obviously conditioned by a worsening of structural perfection with an increase of $x$ and by the corresponding shortening of the free path length, $l$, of a hole photoproduced in the free state (see section 5, item 3). A decrease of $l$ below the distance between adjacent $Mn^{4+}$ ions entails a significant shortening of the hole lifetime and a diminution of photoinduced absorption.

*6.2. Kinetics of photoinduced absorption $\Delta K(t)$*

Figure 3 shows a feature of the kinetics observed under pumping after its switching on: at first, $\Delta K$ rapidly grows (achieving more than half its maximal value in 1 min) and then slowly saturates (the complete saturation not being achieved even after 30 min). This feature can be easily understood in terms of the straggling of potential wells over depth. According to (6), the lifetime of a hole is generally limited by a space $t_{pum} \sim 1$ min sufficient to change a random realization of photoinduced electric fields. It is just the time space $t_{pum}$ during which $\Delta K$ rapidly grows at the beginning of pumping. At $t > t_{pum}$ the growth of $\Delta K$ is slowed down by the recombination of holes released from potential wells by photoinduced fields. Only those holes, which have reached very deep wells, persist and accumulate with time. A low rate of their accumulation and a long delay of the saturation of $\Delta K$ correspond to a small number of very deep wells. A similar feature of the kinetics of $\Delta K$ was observed for $Ca_3Mn_2Ge_3O_{12}$ but it was much more weakly pronounced due to an almost homogeneous structure [13].

The straggling of potential wells over depth manifests itself also in the relaxation of the photoinduced effect after switching off irradiation. It is convenient to characterize the relaxation rate by the ratio

$$\xi = \frac{\Delta K(t_0) - \Delta K(t_0 + t_{rel})}{\Delta K(t_0)} \quad (11)$$

where $t_{rel} = 20$ min is the time of relaxation after the moment, $t_0$, of switching off irradiation. Figure 7 shows $\xi$ versus temperature for $Ca_3Ga_{2-x}Mn_xGe_3O_{12}$ with $x \approx 0.02$ (squares) and $Ca_3Mn_2Ge_3O_{12}$ (circles). In the latter case, $\xi$ rapidly grows with temperature due to a rapid shortening of relaxation time. In the former case, the picture is complicated by the existence of a small number of very deep potential wells. At a high temperature, holes persist only in these wells, whose small number dictates a strong thermal suppression of $\Delta K$ observed under pumping (figure 3(a)). On the other hand, a large depth of persisting populated traps manifests itself in a low relaxation rate $\xi$ shown in figure 7.

Figure 3(b) permits one to trace the manifestation of very deep wells depending on temperature. The figure presents the $\tau$ distribution of decay components derived from the experimental kinetics, $G(\tau)$ (full curves), along with the $\tau$ distribution of holes conditioned by random electric fields (broken curves). At low temperature, the broken and full curves are not strongly different, which means that $G(\tau)$ is formed mainly by random fields. At a high temperature, this is true only for short decay times ($\tau < 300$ s). The maximum of $G(\tau)$ at a longer $\tau$ (of about 800 s) is caused by a small number of holes trapped in very deep wells, which persist at high temperature and provide a long tail of the relaxation curve (figure 3(a)).

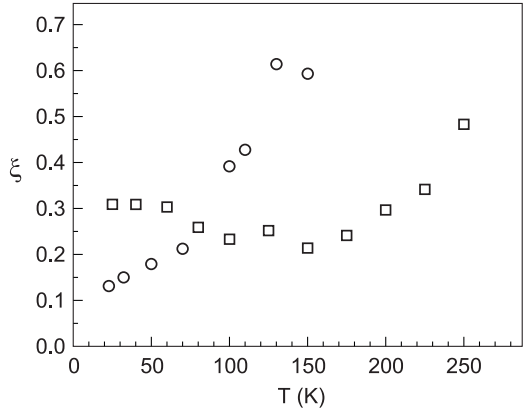

**Figure 7.** Temperature dependence of the relaxation rate (11) observed after switching off irradiation. Squares are related to the sample with $x = 0.02$ ($\lambda_{pum} \approx 510$ nm, $\lambda_{reg} = 375$ nm), circles to the garnet $Ca_3Mn_2Ge_3O_{12}$ ($\lambda_{pum} = 633$ nm, $\lambda_{reg} = 565$ nm) [13].

### 6.3. Effect of the pumping light frequency

A broad distribution of potential wells over depth also entails a nontrivial effect of the pumping light frequency on the temperature dependence of $\Delta K$. Figure 8 shows a significant difference in the character of the dependences $\Delta K(T)$ observed under pumping with green light (photon energy $E \approx 2.43$ eV) and red light ($E = 1.96$ eV). To verify that this difference is not connected with the probe light frequency, in both cases $\Delta K$ was registered at $\lambda_{reg} = 420$ nm (the results are plotted on the left-hand scale); besides, $\Delta K$ produced by green pumping was registered at $\lambda_{reg} = 375$ nm and plotted on the right-hand scale. As seen from figure 8, the temperature dependences of $\Delta K$, observed under green pumping at $\lambda_{reg} = 420$ and $\lambda_{reg} = 375$ nm, practically coincide with each other and essentially differ from the dependence $\Delta K(T)$ under red pumping.

To explain this distinction, an account should be made of the kinetic energy (7) of a photoproduced hole. Judging from the low efficiency of the red pumping, the corresponding photon energy of 1.96 eV slightly exceeds the threshold energy in (7); hence, although the photon energy of the green and red pumping differ only by about 20%, the corresponding values of the kinetic energy (7) differ by much more. This entails a significant difference in the path length $l$ covered by the photoproduced hole before going into a polaronic state (see section 5, item 3). Under pumping with red light, due to a small $l$, a photoproduced hole immediately recombines unless it is trapped by a very deep potential well. A small number of such deep wells conditions a small addition to absorption produced by red pumping. But holes trapped in these deep wells are practically insensitive to an increase of temperature up to 120 K; this is mirrored by a horizontal section of the curve $\Delta K = \Delta K(T)$.

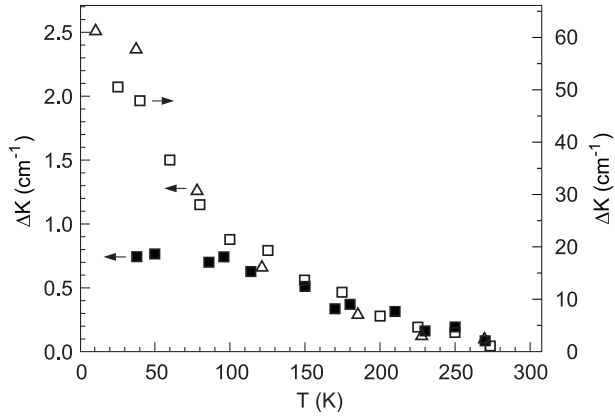

**Figure 8.** Effect of the pumping wavelength on the temperature dependence of photoinduced absorption for the sample with $x = 0.02$. Full squares show the irradiation with red light ($\lambda_{pum} = 633$ nm, $\lambda_{reg} = 420$ nm). Open squares and triangles show the irradiation with green light ($\lambda_{pum} \approx 510$, $\lambda_{reg} = 375$ and 420 nm, respectively).

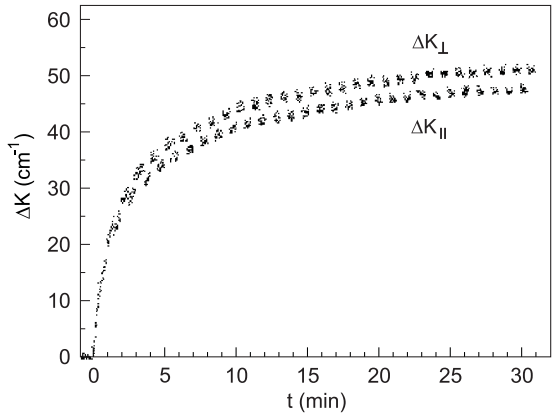

**Figure 9.** Time dependences of the photoinduced additions to absorption registered in the polarizations perpendicular ($\Delta K_\perp$) and parallel ($\Delta K_\parallel$) to the pumping light polarization. The measurements were carried out for $x = 0.02$ at $\lambda_{reg} = 375$ nm, $\lambda_{pum} \approx 510$ nm, $T = 32$ K under full pumping switched on at $t = 0$.

## 7. Photoinduced dichroism

Figure 9 shows the time dependence of photoinduced absorption under pumping by linearly polarized light, the probe light being polarized parallel ($\Delta K_\parallel$) or perpendicular ($\Delta K_\perp$) to the pumping polarization. Dichroism, described by the difference $\Delta K_\perp - \Delta K_\parallel$ at a low temperature is 12 times weaker than photoinduced absorption.

In figure 10, the saturation value of dichroism (taken from figure 9) is shown versus temperature by squares for $x = 0.02$. For comparison, a similar dependence, related to $Ca_3Mn_2Ge_3O_{12}$ [13], is shown by circles; it has the same character with about a five times higher ordinate value.

Two questions arise: what is the nature of photoinduced dichroism and why does its temperature dependence differ from that of photoinduced absorption (shown in figure 10 by the full curve)?

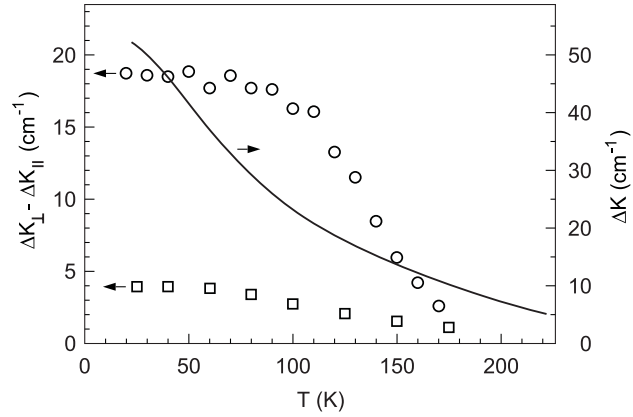

**Figure 10.** Squares: temperature dependence of photoinduced dichroism measured for the sample with $x = 0.02$ at $\lambda_{reg} = 375$ nm under full pumping with $\lambda_{pum} \approx 510$ nm. For comparison, the case of $Ca_3Mn_2Ge_3O_{12}$ [13] is presented by circles. The full curve shows the smoothed temperature dependence of photoinduced absorption at $\lambda_{pum} \approx 510$ nm, $\lambda_{reg} = 375$ nm taken from figure 8.

Photoinduced dichroism should be associated with the anisotropy of two-site hole polarons in the oxygen sublattice. Most of them, as was mentioned in section 5, are trapped in the nearest vicinity of $Mn^{4+}$ ions serving as probe centres. In figure 6, a two-site hole stretched along the $x$ axis is shown in the first coordination sphere of a $Mn^{4+}$ ion. Such holes are preferably generated by pumping light polarized in the $x$ direction. The electric field, produced by this hole at the centre of the $Mn^{4+}$ ion, is oriented parallel to the $z$ axis. The corresponding enhancement of absorption is most pronounced if the probe light is also polarized in the $z$ direction. This results, in agreement with experiment, in a positive difference $\Delta K_\perp - \Delta K_\parallel$.

Now let us consider the temperature dependence of photoinduced dichroism and compare it with that of photoinduced absorption. To that end, it is necessary to trace the processes, leading to the depolarization of an oxygen hole, throughout all stages of its photoproduction and existence listed in section 5. The first stage (excitation of a Mn ion with polarized light) is a pure electronic process independent of temperature.

Attention should be drawn to the second stage—transformation of the atomic excitation of a $Mn^{4+}$ ion to an oxygen hole (a hole analogue of autoionization). This autoionization process, accompanied by overcoming an energy barrier, has in the lattice a thermally assisted character. Hence, an increase of temperature significantly shortens the time, $\tau_{exc}$, required for an excited $Mn^{4+}$ ion to expend its excitation energy in creating an oxygen hole. During the time space $\tau_{exc}$, this atomic excitation (related to the 3d shell partly populated) can be depolarized by a random electric field; such an intra-atomic process cannot strongly depend on temperature. It follows that depolarization on the second stage *diminishes* as $\tau_{exc}$ decreases due to an increase of temperature. In the range of low temperatures, where a large $\tau_{exc}$ makes the second stage essential, this can compensate for an obvious temperature enhancement of the hole depolarization rate on subsequent stages. Taking into account that $\tau_{exc}$ affects only the polarization of a hole but not the event of its formation, the aforesaid explains a different character of the temperature dependences of dichroism and photoinduced absorption (figure 10).

As was shown earlier by the authors [13], dichroism and birefringence [27, 28], induced by linearly polarized light in the $Ca_3Mn_2Ge_3O_{12}$ garnet, are of the same nature, judging from the coincidence of their temperature dependences under pumping and relaxation kinetics after switching off irradiation. Most probably, the mechanism of both phenomena involves two-site hole polarons in the $O^{2-}$ sublattice. The dependence of dichroism and birefringence on the pumping polarization direction is described by phenomenological considerations of symmetry [29], irrespective of the microscopic mechanism.

## 8. Conclusion on the nature of long-lived photoproduced charges and the method of their creation and relaxation

Let us concisely state the main qualitative conclusions drawn from experimental data for the entire set of garnets $Ca_3Ga_{2-x}Mn_xGe_3O_{12}$.

(1) Long-lived photoinduced absorption in the garnets is caused by photoproduced localized charges; a strong degree of localization dictates their long lifetime determined by the rate of their hopping motion resulting in recombination. An alternative mechanism of long-lived changes through photoproduction of lattice defects should be discarded since it is not restricted to a low temperature region and the recovery of photoproduced defects requires at least several days.

(2) Photoproduced long-lived charges should be identified with holes in the $O^{2-}$ sublattice. Indeed, when creating a pair of opposite charges by a photon with energy of 2.5 eV, an electron can be taken away only from an $O^{2-}$ ion, which requires an energy input near 2 eV. The rest of the lattice ions (cations) have ionization potentials of more than 50 eV, much greater than the photon energy. Note that photoproduced holes in the anion sublattice were observed in many ionic crystals (e.g. alkali halides). The motion of photoproduced charges in a regular sublattice is evidenced also by their relaxation kinetics (item 3).

(3) The relaxation kinetics of the photoinduced absorption $\Delta K$ after switching off illumination consists of a continuous set of components with decay times varying in the range of a few orders of magnitude. Such a broad set of decay times corresponds to a broad magnitude distribution (4) of random photoinduced fields which strongly affect the rate of thermally assisted hopping of photoproduced holes and, hence, the rate $\Gamma$ of their relaxation through recombination. Holes with a large $\Gamma$ form a steep initial part of the relaxation curve (shown in figure 3(a)), while charges with a small $\Gamma$ are responsible for its long-time tail.

(4) The mechanism developed of photoinduced absorption $\Delta K$, involving random fields of photoproduced charges, explains a weak dependence of $\Delta K$ on pumping intensity $I$ (the saturation value of $\Delta K$ grows about twice as $I$ increases 20 times). The point is that the actual lifetime of charges under pumping, $\tau(I)$, essentially differs from their lifetime, $\tau = \tau(0)$, in the absence of pumping. If one traces the destiny of a given photoproduced hole, its hopping rate varies by several orders of magnitude during a time, $t_{pum} \approx \text{constant}/I_{pum} \sim 1$ min, sufficient to replace most of the existing charges by new charges photoproduced at random positions. Hence, the saturation value of the number of charges, proportional to the product $I_{pum}\tau(I_{pum})$, is almost independent of $I_{pum}$.

A kinetic equation, making allowance for the statistical distribution of random fields, describes the relaxation kinetics and its observed temperature evolution [13].

(5) A weak dichroism, induced by linearly polarized pumping, is conditioned by a sharp anisotropy of hole polarons in the sublattice of $O^{2-}$ ions with the closed-shell configuration of Ne. By analogy with rare-gas and alkali halide crystals, where the structure of hole polarons was reliably established, a stable hole polaron in the oxygen sublattice must have the form of a two-atom quasi-molecule. It is preferably oriented parallel to the pumping polarization and most closely approaches a probe Mn ion in the perpendicular direction (figure 6). Thus, $\Delta K$ is maximal if measured in the perpendicular polarization, i.e. $\Delta K_\perp > \Delta K_\parallel$, in agreement with experiment.

(6) The distinction in the structure of the garnets $Ca_3Mn_2Ge_3O_{12}$ and $Ca_3Ga_{2-x}Mn_xGe_3O_{12}$ ($x < 0.06$) entails the following distinctions in photoinduced processes:

(a) Although photoinduced addition to absorption in all cases is caused by the electric field of photoproduced charges acting on probe optical centres, the probe centres and the way of this action are different for $Ca_3Mn_2Ge_3O_{12}$ and $Ca_3Ga_{2-x}Mn_xGe_3O_{12}$. The $Ca_3Mn_2Ge_3O_{12}$ garnet has the sublattice of $Mn^{3+}$ ions which serve as probe optical centres. Their green absorption band corresponds to the forbidden transition between even states $^5E_g$ and $^5T_{2g}$ and is enhanced by an electric field that partially eliminates the forbiddenness. In the case of the $Ca_3Ga_{2-x}Mn_xGe_3O_{12}$ garnet, a noticeable photoinduced effect is observed within a violet absorption band that relates to $Mn^{4+}$ ions and corresponds to an allowed transition accompanied by electron transfer between the $O^{2-}$ and $Mn^{4+}$ ions. An applied electric field reduces the energy input, required for charge transfer, and correspondingly shifts the absorption band to the red, thus enhancing absorption at a fixed frequency.

(b) Most probably, photoproduced charges in $Ca_3Mn_2Ge_3O_{12}$ and $Ca_3Ga_{2-x}Mn_xGe_3O_{12}$ are of the same nature and should be identified with two-site hole polarons in the oxygen sublattice (see item 2). However, in the case of the $Ca_3Mn_2Ge_3O_{12}$ garnet with the $Mn^{3+}$ sublattice and a small $Mn^{4+}$ impurity, another mechanism is also possible [12, 13]: the excitation of the Mn sublattice releases part of the Mn holes (pinned at negative Ge vacancies in the ground state) and makes them move over the $Mn^{3+}$ sublattice. The former version, common for $Ca_3Ga_{2-x}Mn_xGe_3O_{12}$ and $Ca_3Mn_2Ge_3O_{12}$, naturally explains a similar character of all photoinduced phenomena observed; however, this is not sufficient grounds to finally discard the latter version for $Ca_3Mn_2Ge_3O_{12}$.

(c) The structure inhomogeneity of the $Ca_3Ga_{2-x}Mn_xGe_3O_{12}$ garnet, caused by Mn impurity, entails a large straggling of potential wells, serving as traps for hole polarons, over depth. Photoproduced holes, trapped in very deep wells, are stable and responsible for the following features typical only of $Ca_3Ga_{2-x}Mn_xGe_3O_{12}$ (see section 6, item 2). First, a small number of very deep wells is slowly populated under pumping, entailing a long delay of the complete saturation of photoinduced absorption $\Delta K$. Second, at high temperatures only holes with a long lifetime, trapped in very deep wells, persist and manifest themselves in a long relaxation of $\Delta K$ after switching off irradiation (the relaxation time does not diminish as temperature increases to 150 K). Holes, trapped in deep wells, are also selected in a similar way when pumping with red light; this manifests itself in a slow temperature dependence of $\Delta K$ under red pumping. Third, a worsening of the lattice perfection with an increase of the manganese content $x$ at $x > 0.02$ results in a decrease of the distance of a photoproduced hole from its recombination centre and in a diminution of the photoinduced effect.